\documentclass[
prd,
a4paper,
nofootinbib,
showpacs
]{revtex4}
\usepackage{graphicx}
\usepackage{amsmath}
\usepackage{amssymb}
\def\Tr{\mbox{Tr}\,}
\newcommand{\vev}[1]{\langle #1\rangle}
\begin{document}
\title{The quantum vacuum of the minimal SO(10) GUT}

\author{Stefano Bertolini}\email{bertolin@sissa.it}
\author{Luca Di Luzio}\email{diluzio@sissa.it}
\affiliation{INFN, Sezione di Trieste, and SISSA,
via Bonomea 265, 34136 Trieste, Italy}
\author{Michal Malinsk\'y}
\email{malinsky@kth.se}
\affiliation{Theoretical Particle Physics Group,
Department of Theoretical Physics,
Royal Institute of Technology (KTH),
Roslagstullsbacken 21,
SE-106 91 Stockholm, Sweden}
\begin{abstract}
We reexamine the longstanding no-go excluding all potentially viable $SO(10)\to SU(3)_{c}\otimes SU(2)_{L}\otimes U(1)_{Y}$ symmetry breaking patterns within the minimal renormalizable non-supersymmetric SO(10) GUT framework featuring the $45$-dimensional adjoint representation in the Higgs sector. A simple symmetry argument indicates that quantum effects do change the vacuum structure of the model dramatically. A thorough analysis of the one-loop effective potential reveals that the phenomenologically favoured symmetry breaking chains passing through the $SU(4)_C \otimes SU(2)_L \otimes U(1)_R$ or
$SU(3)_c \otimes SU(2)_L \otimes SU(2)_R \otimes U(1)_{B-L}$ intermediate stages 
are, indeed, supported at the quantum level. This brings the class of minimal non-supersymmetric $SO(10)$ GUTs back from oblivion, providing a new ground for a potentially realistic model building.
\end{abstract}
\pacs{12.10.Dm, 11.15.Ex, 11.30.Qc}
\maketitle
\section{Introduction}
In spite of the enormous progress the field of grand unified theories (GUTs) experienced since the seminal works of Georgi and Glashow \cite{Georgi:1974sy} it is still not clear whether the basic scheme is by any means reflected by nature. Nevertheless, the inherent testability of the simplest GUT scenarios makes the idea of a unified description of the strong and electroweak interactions so popular that there has been a lot of efforts spent recently in attempts to pin down the simplest potentially viable scenarios, in particular those based on the $SU(5)$ \cite{Bajc:2002pg} and/or $SO(10)$ \cite{Albright:2000dk,Aulakh:2003kg} gauge symmetry. Apart from the canonical test imposed by the required high-energy convergence of all three Standard Model (SM) gauge couplings, a successful candidate should obey also the constraints inflicted by the proton longevity (stretching over $10^{33}$ years) and the absolute neutrino mass scale, which are indeed closely related to the generic need for a compatibility with the observed SM flavour structure.    

However, the situation is not very clear at the moment as even the very concept of minimality admits a certain number of interpretations. For instance, it can be understood as a mere ``simplicity'' related to e.g. a minimum rank of the gauge group, the Higgs sector dimensionality, the naturalness of the doublet-triplet splitting, or even the complexity of the gauge unification pattern. In this respect, the models based on $SU(5)$ gauge group are frequently regarded to as more minimal than the $SO(10)$ ones. Alternatively, predictivity (i.e., the number of independent couplings) is often invoked as a measure of minimality; what matters in such a case is namely the renormalizability of the theory, the number of Higgs multiplets, the complexity of the GUT-scale Yukawa sector, the presence or absence of supersymmetry (SUSY) etc. In this respect, the $SO(10)$ GUTs (especially their SUSY variants) usually score better than the $SU(5)$ models, see e.g. \cite{Aulakh:2003kg,Fukuyama:2004pb} and references therein. To us, even though potentially troublesome on the naturalness side, this is a very physical option - besides giving generally better prospects for an ultimate testability the preference of $SO(10)$ also offers a natural relief from the persistent trouble with the simplest $SU(5)$ models, failing either on the unification side (like e.g. the original Georgi-Glashow model) or suffering from an overly fast proton decay when SUSY is invoked\footnote{For a nice account for the proton decay issue in the minimal SUSY $SU(5)$ see e.g. \cite{Bajc:2002pg}.}. 

Sticking to the $SO(10)$ case, minimality is essentially equivalent to the complexity of the Higgs sector. In the class of $SO(10)$ GUTs, the choice of the Higgs representation is subject to several basic requirements. First, there is the need to break consistently the rank=5 $SO(10)$ gauge symmetry down to the rank=4 $SU(3)_{c}\otimes SU(2)_{L}\otimes U(1)_{Y}$ of the SM. Remarkably, none of the basic candidate Higgs representations (namely, ${\bf 16}$ and/or  ${\bf 126}$) capable of breaking the $U(1)_{B-L}$ subgroup 
of $SO(10)$, can account for the full $SO(10)\to {\rm SM}$  breakdown on its own, since their SM singlets leave a full $SU(5)$ subgroup of  $SO(10)$ intact. From the group theory point of view, the simplest solution one can devise is an extra adjoint ${\bf 45}$ in the Higgs sector whose additional pair of VEVs brings in enough freedom to provide the full $SO(10)\to  SU(3)_{c}\otimes SU(2)_{L}\otimes U(1)_{Y}$ breakdown. Unfortunately, this approach fails in the SUSY case where the extra $F$-flatness conditions force the VEVs in ${\bf 45}$ to align along the direction of $\vev{\bf 16}$ and thus $SU(5)$ still remains intact \cite{Aulakh:2000sn}. Thus, $\bf 210$ is usually invoked in the SUSY $SO(10)$ context, giving rise to the minimal SUSY $SO(10)$ GUT with a Higgs sector spanning over the reducible representation ${\bf 10}\oplus \overline{{\bf 126}}\oplus {\bf 126}\oplus {\bf 210}$, see e.g. \cite{Clark:1982ai,Aulakh:1982sw} and also \cite{Aulakh:2003kg}. 
   
However, though successful in accounting for some of the basic features of the lepton flavour mixing \cite{Bajc:2002iw}, the minimal renormalizable SUSY SO(10) model fails due to a generic tension between the absolute neutrino mass scale and the gauge coupling unification constraints, \cite{Aulakh:2005mw,Bertolini:2006pe} and there is a general need to go beyond the simplest setting, see e.g. \cite{Bertolini:2004eq,Aulakh:2006hs,Bajc:2008dc,Melfo:2010gf} and references therein.

On the other hand, if predictivity rather than naturalness is at stakes, supersymmetry is not a mandatory ingredient of the $SO(10)$ GUTs because the unification constraints refuting the simplest non-SUSY $SU(5)$ settings can be satisfied with an intermediate energy scale $M_{I}$ below the scale of grand unification $M_{G}$.  In particular, the recent analysis of the relevant symmetry breaking chains \cite{Bertolini:2009qj} reveals an interesting picture associated to the two basic single-step chains 
\begin{eqnarray}
\label{chainXII}
SO(10)&\stackrel{M_G}{\longrightarrow}
&SU(4)_C \otimes SU(2)_L \otimes U(1)_R \; \stackrel{M_I}{\longrightarrow} \; \mbox{SM}\,,\\
\label{chainVIII}
SO(10)&\stackrel{M_G}{\longrightarrow}
& SU(3)_c \otimes SU(2)_L \otimes SU(2)_R \otimes U(1)_{B-L} \; \stackrel{M_I}{\longrightarrow} \;  \mbox{SM}\,,
\end{eqnarray}
driven by the simplest conceivable Higgs sector  spanned over just ${\bf 45}\oplus {\bf 16}$ of $SO(10)$. Unfortunately, this scenario, requiring in practice a few orders of magnitude hierarchy between $M_{I}$ and $M_{G}$, has been long ago \cite{Yasue,Anastaze:1983zk,Babu:1984mz} claimed incompatible with the dynamics of the scalar sector. The reason is an emergence of tachyonic masses in the Higgs spectrum unless $M_{I}$ essentially coincides with $M_{G}$, at odds with the running constraints. 

All these arguments are, however,  based on the tree-level analysis of the Higgs potential. Remarkably enough, the recent work \cite{Bertolini:2009es} reveals a dramatic qualitative change of the picture once quantum effects are properly taken into account, with a potential of bringing the class of minimal non-supersymmetric $SO(10)$ GUTs back to life. 
\section{The classical failure of the minimal non-SUSY $SO(10)$ GUT - a closer look}
In order to fully appreciate this fact one has to inspect in detail the available vacua of the renormalizable non-SUSY scalar potential spanned over ${\bf 45}\oplus {\bf 16}$ in the Higgs sector. However, before entering the realm of quantum effects one has to understand the origin of the tree-level no-go implied by \cite{Yasue,Anastaze:1983zk,Babu:1984mz}.

The most generic classical renormalizable scalar potential one can write with only ${\bf 45}\oplus {\bf 16}$ at hand reads $V_0=V_{45}+V_{16}+V_{\rm int}$, where \cite{Buccella:1980qb}
\begin{eqnarray}
\label{potentialV45}
V_{45}&=&
-\frac{\mu^2}{2}\Tr{\bf 45}^2 + \frac{a_1}{4}(\Tr{\bf 45}^2)^2 + \frac{a_2}{4}\Tr{\bf 45}^4 \, , \\
\label{potentialV16}
V_{16}&=&
-\frac{\nu^2}{2}{\bf 16}^\dag{\bf 16}
+\frac{\lambda_1}{4}({\bf 16}^\dag{\bf 16})^2
 +\frac{\lambda_2}{4}({\bf 16}_+^\dag\Gamma_j{\bf 16}_-)({\bf 16}_-^\dag\Gamma_j{\bf 16}_+) \,
 \end{eqnarray}
correspond to the mass and self-interaction terms for $\bf 45$ and $\bf 16$, respectively, while
\begin{eqnarray}
\label{potentialV4516}
V_{\rm int}=
\alpha({\bf 16}^\dag{\bf 16})\Tr{\bf 45}^2+\beta{\bf 16}^\dag{\bf 45}^2{\bf 16}
+\tau{\bf 16}^\dag{\bf 45}\,{\bf 16} \,
\end{eqnarray}
provides the desired interaction between the two sectors.
Employing an $SU(3)_c \otimes SU(2)_L \otimes SU(2)_R \otimes U(1)_{B-L}$ notation the vacuum manifold is parametrized by the VEVs of the  three different SM singlets   
\begin{equation}
\label{vevs}
\vev{(1,1,1,0)_{45}}\equiv \omega_{Y}\,,\;\;
\vev{(1,1,3,0)_{45}}\equiv \omega_{R}\,,\;\;
\vev{(1,1,2,+1/2)_{16}}\equiv \chi_{B-L}\,.
\end{equation}
Since the gauge unification in this minimal scenario typically requires the $B-L$ breaking scale $\chi_{B-L}$ to be lower than the higher from the pair $\omega_{R}$, $\omega_{Y}$ (to be identified with $M_{G}$),  the heavy Higgs spectrum is governed namely by the VEVs of $\bf 45$. Apart from the generic setting with non-zero $\omega_{R}\neq \omega_{Y}$ corresponding to an intermediate $SU(3)_c \otimes SU(2)_L \otimes U(1)_R \otimes U(1)_{B-L}$ stage, there are several important limiting cases exhibiting an enhanced intermediate symmetry, in particular
\begin{eqnarray}
\label{vacua}
\omega_{R}\neq 0,\, \omega_{Y}= 0\; &: & SU(4)_C \otimes SU(2)_L \otimes U(1)_R\,,  \\
\omega_{R}= 0,\, \omega_{Y}\neq 0\;  &: &  SU(3)_c \otimes SU(2)_L \otimes SU(2)_R \otimes U(1)_{B-L}\,, \nonumber \\
\omega_{R}=\pm \omega_{Y}\neq 0\; &:  & \mbox{standard ($+$) or flipped ($-$)}\; SU(5)\otimes U(1)\nonumber \,.
\end{eqnarray}
The point is that while the first two cases (corresponding to the simplest conceivable chains (\ref{chainXII}), (\ref{chainVIII})) are the only ones from list (\ref{vacua}) which are compatible with the non-SUSY unification constraints \cite{Bertolini:2009es}, they seem to be disfavoured by the detailed Higgs sector dynamics \cite{Yasue,Anastaze:1983zk,Babu:1984mz} due to the emergence of tachyons within the corresponding vacua. 
\subsection{The emergence of tachyons in the non-$SU(5)\otimes U(1)$ vacua} 
This is due to the fact that a full-fledged analysis of the classical Higgs potential (\ref{potentialV45})-(\ref{potentialV4516}) reveals a pair of scalar multiplets with the classical SM-level masses given by the formulae 
\begin{eqnarray}
\label{PGBmasses}
M^2(1,3,0)_{45} & = &
2 a_2 (\omega _Y - \omega _R) (\omega _Y + 2 \omega _R) \, , \\
M^2(8,1,0)_{45} & = &
2 a_2 (\omega _R - \omega _Y) (\omega _R + 2 \omega _Y) \,.\nonumber
\end{eqnarray}
It is clear that in the majority of the parametric space, in particular, everywhere outside the domain 
\begin{equation}
\label{classicallyallowed}
-2\leq\omega_{Y}/\omega_{R}\leq-1/2\,,
\end{equation}
 at least one of them develops a negative mass-squared, indicating an instability of the corresponding vacuum configuration. Consequently, all stable settings exhibit an intermediate flipped $SU(5)\otimes U(1)$ stage, at odds with unification constraints. This, indeed,  is the core of the classical argument against the potential viability of the minimal non-SUSY $SO(10)$ GUT framework.   
  
Since the simplicity of the formulae (\ref{PGBmasses}) is crucial for the above reasoning it is important to understand their structure in as much detail as possible. In this respect, it is particularly illuminating to discuss the global symmetries of the scalar potential $V_{0}$. 

\subsection{Understanding the structure of the key mass formulae  in terms of symmetries}
When only trivial invariants (i.e. moduli) of both ${\bf 45}$ and ${\bf 16}$ are considered,
the global symmetry of $V_{0}$
is $O(45)\otimes O(32)$.
It is spontaneously broken down to $O(44)\otimes O(31)$ by the ${\bf 45}$ and ${\bf 16}$ VEVs
yielding 44+31=75 Goldstone bosons in the scalar spectrum.
Since in this case also the gauge $SO(10)$ symmetry is broken to the SM gauge group,
$45-12=33$ would-be Goldstone bosons, with the quantum numbers
of the coset $SO(10)/{\rm SM}$ algebra,
decouple from the physical spectrum,
while $75-33=42$ pseudo-Goldstone bosons remain.
Their masses are generally expected to receive contributions
from the explicitly breaking terms $a_2$, $\lambda_2$, $\beta$ and $\tau$.

However, the tree-level masses of the crucial $(1,3,0)_{45}$ and $(8,1,0)_{45}$ multiplets, which belong among these pseudo-Goldstone bosons, turn out to depend only on the parameter $a_2$ but {\em not} on the other parameters expected, 
namely $\lambda_2$, $\beta$ and $\tau$.
While the $\lambda_2$ and $\tau$ terms cannot obviously contribute
at the tree level to ${\bf 45}$ mass terms,
one would generally expect a contribution from the $\beta$ term proportional to $\chi_{B-L}^2$ to pop up in relations~(\ref{PGBmasses}).
This, however, is not the case.
One can understand that on general grounds
by observing that
the relevant scalar interaction in (\ref{potentialV4516}) resembles the
the covariant-derivative interaction responsible for the contribution of $\vev{\bf 16}$ to the gauge boson masses.
As a consequence, no tree-level mass contribution from the $\beta$ coupling can be generated  for the scalars in ${\bf 45}$
which carry the SM quantum numbers of the $W^{\pm}$ and $Z^{0}$ bosons, i.e., $(1,3,0)_{45}$, and gluons $(8,1,0)_{45}$ that are protected by the residual $SU(5)$ symmetry left unbroken even with $\vev{{\bf 16}}\neq 0$.

\section{The quantum structure of the minimal non-SUSY $SO(10)$ GUT vacuum}
On the other hand, we should expect the $\tau$ and $\beta$ interactions to contribute to the masses of $(1,3,0)_{45}$ and $(8,1,0)_{45}$ at the quantum level. While for the former the loops simply admit to contract the two ${\bf 16}$'s sticking out of the $\tau$-vertex in (\ref{potentialV4516}), a similar loop due to a pair of $\beta$-vertices brings into play the $SU(5)$-breaking VEVs of $\bf 45$.  
Similar contributions should also arise from the gauge interactions
which break explicitly the independent global transformations on
the ${\bf 45}$ and ${\bf 16}$ discussed in the previous section.

\begin{figure}[h]
\includegraphics[width=12pc]{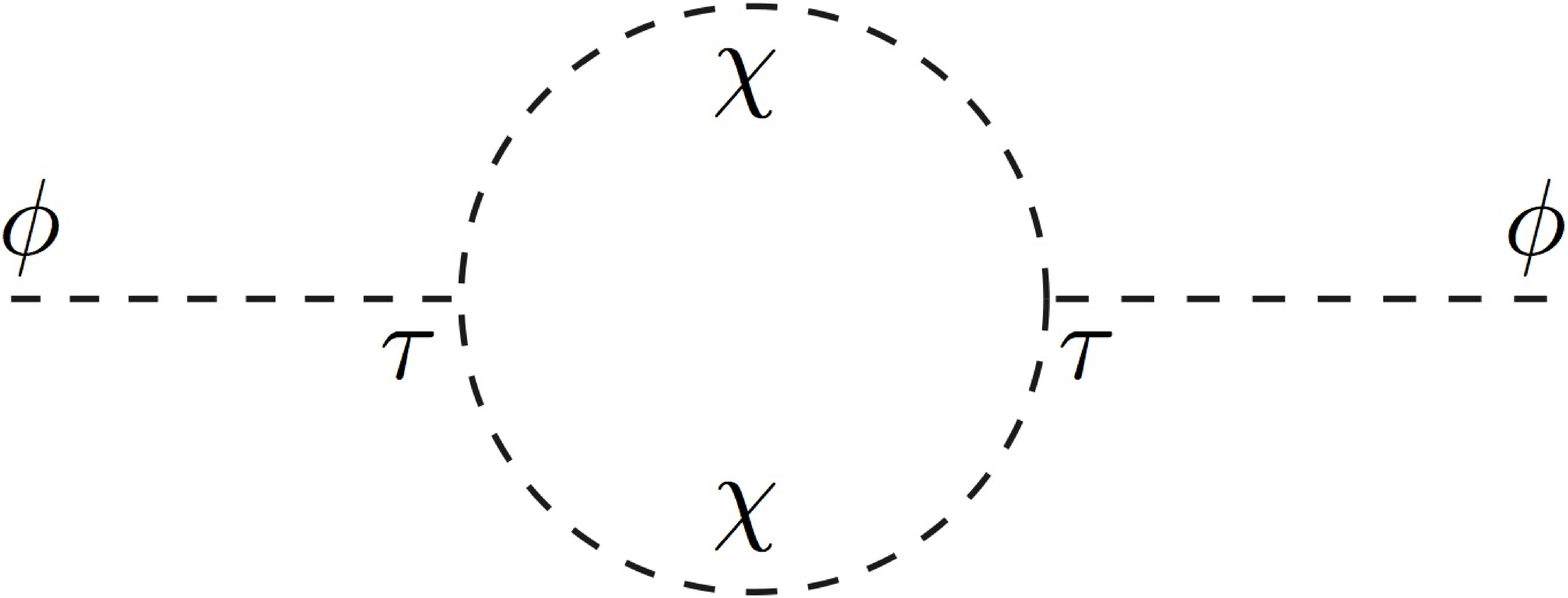}\quad
\includegraphics[width=12pc]{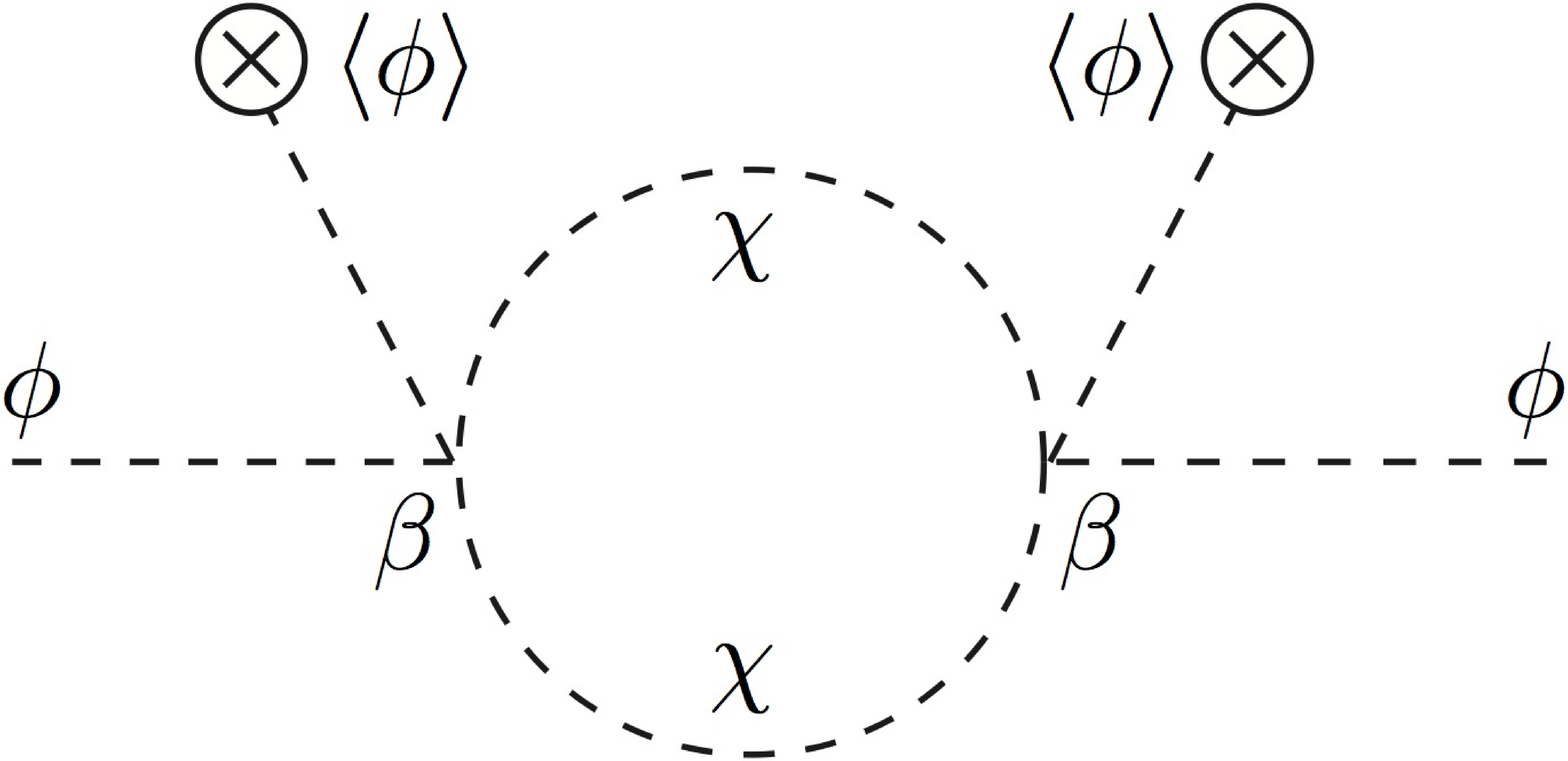}\quad
\includegraphics[width=12pc]{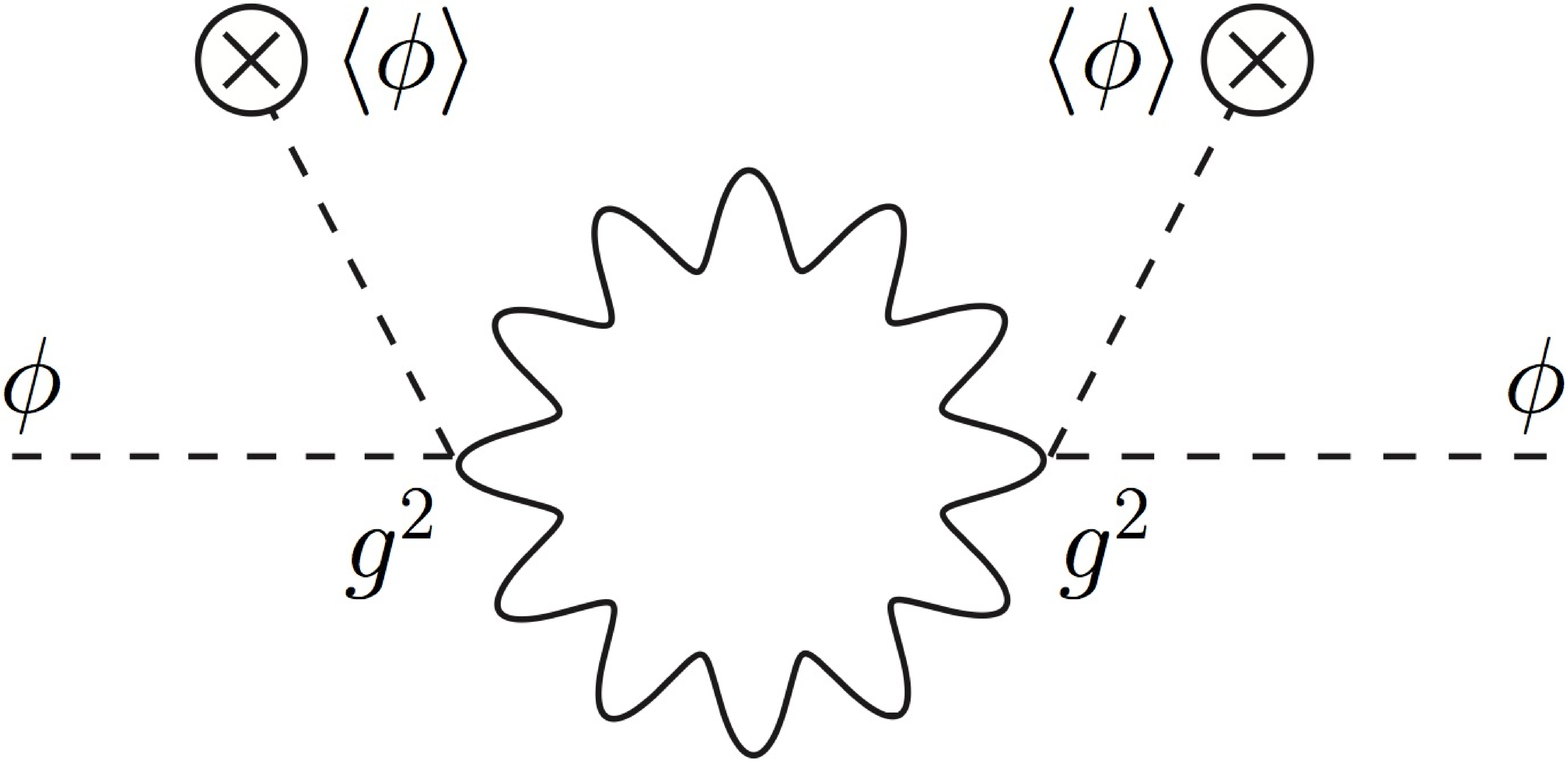}
\caption{\label{graphs}The diagrammatic structure of the three important types of quantum corrections which stabilise the vacua supporting the desired $SU(4)_C \otimes SU(2)_L \otimes U(1)_R$ (case A) or $SU(3)_c \otimes SU(2)_L \otimes SU(2)_R \otimes U(1)_{B-L}$ (case B) symmetry breaking chains at the one-loop level. The dangerous tachyonic mass parameters of the $(1,3,0)_{45}$ and/or $(8,1,0)_{45}$ Higgs multiplets in ${\bf 45}$ are lifted provided ${8\pi^{2}}|a_{2}|<\tau^{2}/\omega_{R}^{2}+\beta^{2}+13g^{4}$ in case A, or  ${8\pi^{2}}|a_{2}|<\tau^{2}/\omega_{Y}^{2}+2\beta^{2}+19g^{4}$ in case B. The interested reader can find further details in the original work \cite{Bertolini:2009es}.}
\end{figure}

In the diagrammatic language, the relevant one-loop self-energies are depicted in Figure~\ref{graphs} where the individual components of the scalars transforming as {\bf 45} and {\bf 16} of $SO(10)$ have been denoted by generic symbols $\phi$ and $\chi$, respectively.
While the exchange of the components of ${\bf 16}$ is crucial,
the $\chi_{B-L}$ does not need to enter;
in the phenomenologically allowed unification patterns it gives actually negligible contributions.


One is thus lead to the conclusion that any result based on the particular shape of the tree-level ${\bf 45}$ vacuum is  dramatically altered at the quantum level. Let us emphasise that although one may in principle avoid the $\tau$-term by means of e.g. an extra $Z_{2}$ symmetry, no symmetry can forbid the $\beta$-term and the gauge loop contributions.
\subsection{The one-loop effective potential analysis}
The quantum structure of the model vacuum is best assessed by the minimisation of the effective potential \cite{Coleman:1973jx}. It can be conveniently written as $V_{\rm eff}=V_{0}+\Delta V$ where the leading one-loop contribution to $\Delta V$ can be cast in the form
\begin{equation}\label{DeltaV}
\Delta V(\phi ,\chi ,\mu) =
\frac{1}{64\pi^2}
\Tr\left[\mathcal{M}_{s}^4(\phi ,\chi)\left(\log\frac{\mathcal{M}_{s}^2(\phi ,\chi)}{\mu^2}
-\frac{3}{2}\right)+3\mathcal{M}_{g}^4(\phi ,\chi)\left(\log\frac{\mathcal{M}_{g}^2(\phi ,\chi)}{\mu^2}
-\frac{5}{6}\right)\right] \, .
\end{equation}
Here the terms in the bracket correspond to the contribution from the (real) scalars and gauge bosons, respectively, and $\mathcal{M}_{s}$ and $\mathcal{M}_{g}$ denote the relevant field-dependent tree-level mass matrices. Note also that the contribution due to the SM fermions has been neglected because of the hierarchical suppression of their masses with respect to the GUT-scale Higgs and gauge bosons. For completeness, let us also mention that the specific form of $\Delta V$  corresponds to the Landau gauge and the $\overline{\rm MS}$ renormalization scheme.

Using a generic symbol $\psi$ for any of the components of ${\bf 45}\oplus {\bf 16}$ the first derivative of the one-loop part of the effective potential (\ref{DeltaV}) is given by
\begin{equation}
\label{der1deltaV}
\frac{\partial\Delta V}{\partial\psi}=
\frac{1}{64\pi^2}\Tr\left[\left\{\frac{\partial\mathcal{M}_{s}^2}{\partial\psi},
  \mathcal{M}_{s}^2\right\}  
\left(\log\frac{\mathcal{M}_{s}^2}{\mu^2}-\frac{3}{2}\right)
+ \mathcal{M}_{s}^2\frac{\partial\mathcal{M}_{s}^2}{\partial\psi}\right]+{\rm gauge\;part} \,,
\end{equation}
where, for the sake of simplicity, only the scalar contribution due to the first term in eq.~(\ref{DeltaV}) has been written explicitly; the gauge part derived from the $\mathcal{M}_{g}$-piece of (\ref{DeltaV}) is completely analogous. Let us also note that the formula above holds regardless of the commutation properties of $\mathcal{M}$ and its first derivatives. 

Similarly, the second derivatives of $\Delta V$ can be written in the form
\begin{eqnarray}
\frac{\partial^2\Delta V}{\partial\psi_a\partial\psi_b}&=&
\frac{1}{64\pi^2} \Tr \Biggl\{\nonumber \\
& \times & \left[\left\{\frac{\partial^{2} \mathcal{M}_{s}^2}{\partial\psi_a\partial\psi_b}, \mathcal{M}_{s}^2\right\}
+ \left\{ \frac{\partial \mathcal{M}_{s}^2}{\partial\psi_a}, \frac{\partial \mathcal{M}_{s}^2}{\partial\psi_b}\right\}\right]
\left(\log\frac{\mathcal{M}_{s}^2}{\mu^2}-\frac{3}{2}\right)
+\frac{\partial \mathcal{M}_{s}^2}{\partial\psi_a}\frac{\partial \mathcal{M}_{s}^2}{\partial\psi_b}
+\mathcal{M}_{s}^2\frac{\partial^{2} \mathcal{M}_{s}^2}{\partial\psi_a\partial\psi_b}\nonumber \\
&+& \sum_{m=1}^{\infty}(-1)^{m+1}\frac{1}{m}\sum_{k=1}^{m}{m\choose k}\left\{ \mathcal{M}_{s}^{2}, \frac{\partial \mathcal{M}_{s}^2}{\partial\psi_a}\right\}  
\left[\mathcal{M}_{s}^{2},..\left[\mathcal{M}_{s}^{2}, \frac{\partial \mathcal{M}_{s}^2}{\partial\psi_b}\right]..\right]\left(\mathcal{M}_{s}^{2}-1\right)^{m-k}
\Biggr\}\nonumber\\
&+&{\rm gauge\;part} \,,
\label{der2deltaV}
\end{eqnarray}
where the infinite series of nested commutators (taken $k-1$ times for each $k$) looks after the general non-commutativity of $\mathcal{M}_{s}$ and $\partial \mathcal{M}_{s}^2/\partial\psi$ which becomes relevant here.

Note also that such terms are also crucial for a consistent cancellation of the infrared divergencies emerging in the one-loop mass matrix $\overline{m}^{2}_{ab}\equiv \partial^{2}V_{\rm eff}/\partial\psi_{a}\partial\psi_{b}$ when evaluated for the solutions of the one-loop stationary equations $\partial V_{\rm eff}/\partial\psi_{a}=0$. These are, of course, due to the presence of Goldstone bosons in the Landau gauge inherent to formula (\ref{DeltaV}).  Let us recall that the effective potential, however, is defined at zero external momenta, and the physical Higgs spectrum is derived from the ``shifted'' mass matrix $M^{2}_{ab}(p^{2})= \overline{m}^{2}_{ab}+\Delta\Sigma_{ab}(p^{2})$ for $p^{2}=M_{a}^{2}$ where $\Delta\Sigma_{ab}(p^{2})\equiv \Sigma_{ab}(p^{2})-\Sigma_{ab}(0)$ is the difference of the relevant one-loop self-energies evaluated at two different four-momenta. 
\subsection{The one-loop corrections to the masses of $(1,3,0)_{45}$ and $(8,1,0)_{45}$}
Putting all this together, the quantum shifts in the masses of the two critical multiplets read
\begin{eqnarray}
\label{310onthevac}
\Delta M^2(1,3,0)_{45}&=& \frac{1}{4\pi^2} \left[ \tau^2
+\beta^2(2\omega_R^2-\omega_R\omega_Y+2\omega_Y^2)
+g^4 \left(16 \omega _R^2+\omega _Y \omega _R+19 \omega _Y^2\right)\right]+{\rm logs}\,, \;\;\; 
\nonumber\\
\label{810onthevac}
\Delta M^2(8,1,0)_{45}&=& \frac{1}{4\pi^2} \left[ \tau^2
+\beta^2(\omega_R^2-\omega_R\omega_Y+3\omega_Y^2)
+g^4 \left(13 \omega _R^2+\omega _Y \omega _R+22 \omega _Y^2\right)\right]+{\rm logs}\,,\nonumber 
\end{eqnarray}
where the symbols ``logs'' denote the sub-leading logarithmic contributions. It is easy to see that with these terms added to the tree-level structure (\ref{PGBmasses}) both $(1,3,0)_{45}$ and $(8,1,0)_{45}$ mass-squares can be made simultaneously positive for wide ranges of the $\omega_{R}$ and $\omega_{Y}$ parameters, well outside the classical domain (\ref{classicallyallowed}). This holds, especially, for the configurations with either $\omega_{Y}\ll \omega_{R}$ or $\omega_{R}\ll \omega_{Y}$ corresponding to the classically forbidden $SU(4)_C \otimes SU(2)_L \otimes U(1)_R$ (case A) or $SU(3)_c \otimes SU(2)_L \otimes SU(2)_R \otimes U(1)_{B-L}$ (case B) intermediate symmetries. In particular, the dangerous tachyonic mass parameters of the $(1,3,0)_{45}$ and/or $(8,1,0)_{45}$ Higgs multiplets in ${\bf 45}$ are lifted for ${8\pi^{2}}|a_{2}|<\tau^{2}/\omega_{R}^{2}+\beta^{2}+13g^{4}$ in case A, or  ${8\pi^{2}}|a_{2}|<\tau^{2}/\omega_{Y}^{2}+2\beta^{2}+19g^{4}$ in case B.

Thus, the quantum effects restore the  full symmetry breaking potential of the minimal non-SUSY $SO(10)$ Higgs sector spanned over ${\bf 45}\oplus {\bf 16}$, bringing back into play the symmetry breaking chains favoured by the gauge coupling unification constraints! Let us also stress that the fact that $U(1)_{B-L}$ has been broken by the spinorial $\bf 16$  does not play a major role in the argument so an alternative Higgs model based on ${\bf 45}\oplus {\bf 126}$ representation exhibits the same qualitative behaviour. Similarly, an extra ${\bf 10}$ added to the Higgs sector in order to get potentially realistic Yukawa couplings makes no harm because it can not develop any GUT-scale VEVs.
\section{Conclusions and outlook}
Typically, only the minimal GUTs can be subject to a thorough scrutiny by complementary methods such as the proton lifetime constraints, the absolute neutrino mass scale or the match between the effective SM flavour structure and the underlying GUT-scale Yukawa sector. In this respect, the recent failure of the minimal renormalizable SUSY $SO(10)$ GUT re-opened the question of the potential viability of the simplest non-SUSY $SO(10)$ scenarios. Remarkably, the generic no-go obtained about 30 years ago, which plagued the simplest model with ${\bf 45}\oplus {\bf 16}$ in the Higgs sector, has been recently identified as a mere artefact of the classical approach that is lifted at the quantum level. This reiterates the intriguing question whether a potentially viable and intrinsically consistent non-SUSY $SO(10)$ theory can be spanned over a renormalizable Higgs sector with only ${\bf 45}\oplus {\bf 16}$ (or, alternatively, ${\bf 45}\oplus {\bf 126}$) participating at the GUT symmetry breakdown. The final answer, however, is left to a future analysis.

\end{document}